\documentclass[aps,prb,preprint,amsmath,amssymb,floatfix,draft]{revtex4}
\usepackage{amsmath}
\usepackage{amsfonts}

\begin{document}

\title{Three fermions with six single particle states can be
entangled in two inequivalent ways}
\author{P\'eter L\'evay and P\'eter Vrana}
\affiliation{Department of Theoretical Physics, Institute of
Physics, Budapest University of Technology and Economics}
\date{\today}

\begin{abstract}
Using a generalization of Cayley's hyperdeterminant as a new
measure of tripartite fermionic entanglement we obtain the SLOCC
classification of three-fermion systems with six single particle
states. A special subclass of such three-fermion systems is shown
to have the same properties as the well-known three-qubit ones.
Our results can be presented in a unified way using Freudenthal
triple systems based on cubic Jordan algebras. For systems with an
arbitrary number of fermions and single particle states we propose
to use the Pl\"ucker relations as a sufficient and necessary
condition of separability.

\end{abstract}
\maketitle{}

\section{Introduction}
The quantification of multipartite entanglement is one of the most
important problems of quantum information theory. Regarding
entanglement as a resource proved to be a useful idea producing
spectacular applications such as teleportation\cite{Jozsa},
cryptography\cite{Bennett} and quantum computing\cite{Nielsen},
and paved the way for further possible fascinating applications.
However, this idea immediately leads us also to the need of
classifying different types of entanglement via suitable
entanglement measures. These measures are real-valued functions of
quantum states trying to quantify the amount of entanglement these
states contain. For
systems of {\it distinguishable} constituents characterized by
either pure or mixed states 
on the structure of such entanglement measures 
a great variety of results is
available\cite{Virmani,Horodecki,Bengtsson}.
 However, much less is known about the structure of multipartite
 entanglement measures for systems with
{\it indistinguishable} constituents. For {\it bipartite}
fermionic and bosonic systems a number of useful results
exists\cite{Annals,Li,Paskauskas,Ghirardi,Sanders,Schlie}. For
example for {\it two}-fermion systems having $2K$ single particle
states a decomposition similar to the Schmidt decomposition was
introduced\cite{Schlie}. This Slater decomposition uses the concept
of {\it Slater rank}, i.e. the number of Slater determinants
occurring in the canonical form of any bipartite fermionic system,
for the quantification of fermionic entanglement. The simplest
nontrivial example occurs for a two fermion system with four
single particle states. Here the states are characterized by {\it
six} complex numbers $P_{12},P_{13},P_{14},P_{23},P_{24},P_{34}$
that can be arranged into a $4\times 4$ antisymmetric matrix
$P_{ab}, a,b=1,2,3,4$. It turns out that states of Slater rank one
are the ones for which the Pl\"ucker relation\cite{LNP}
\begin{equation}
P_{12}P_{34}-P_{13}P_{24}+P_{14}P_{23}=0 \label{plucky}
\end{equation}
\noindent holds. From multilinear algebra it is well-known that
this condition is a sufficient and necessary one for writing
$P_{ab}$ in the form: $P_{ab}=v_aw_b-w_av_b$ for some four
component vectors $v$ and $w$ i.e. in this case $P=v\wedge w$, it
is a Slater determinant. Such fermionic states are called {\it
separable}. When the quantity in Eq. (\ref{plucky}) is different
from zero we have states of Slater rank {\it two}, in this case we
have precisely two terms in the Slater decomposition and the state
is {\it entangled}. A useful measure of bipartite entanglement in
this case is\cite{Schlie}
\begin{equation}
0\leq \eta=8\vert P_{12}P_{34}-P_{13}P_{24}+P_{14}P_{23}\vert\leq
1 \label{eta}
\end{equation}
\noindent Notice that the measure satisfies\cite{LNP}
$\eta^2=64\vert{\rm Det (P)}\vert$ hence it is invariant under the
action of the group $SL(4, {\bf C})$ of the form
\begin{equation}
P_{a_1b_1}\mapsto {G_{a_1}}^{a_2}{G_{b_1}}^{b_2}P_{a_2b_2},\qquad
G\in SL(4, {\bf C}), \label{sloci4}
\end{equation}
\noindent where summation for repeated indices is understood. Such
transformations form a subgroup of SLOCC transformations
(stochastic local operations and classical communication)
introduced by D\"ur et.al.\cite{Dur} The SLOCC group is just the
group of invertible $4\times 4$ complex matrices i.e. $GL(4, {\bf
C})$. Two states are SLOCC equivalent iff there exists a $G\in
GL(4,{\bf C})$ transformation converting one state to the other.
Since under a SLOCC transformation ${\eta}\mapsto \vert{\rm
Det}(G)\vert\eta $, there are only {\it two} SLOCC classes
corresponding to the cases $\eta\neq 0$ and $\eta =0$. Since $\eta
=0$ characterizes the separable states there is only one
nontrivial SLOCC class for two fermions with four single particle
states.

As the first nontrivial case of multipartite fermionic
entanglement in this paper we address the classification of the
{\it simplest} of three-fermion systems. By virtue of duality of
forms $\bigwedge^3{\bf C}^4\simeq \bigwedge^1{\bf C}^4$
three-fermion systems with four single particle states can be
mapped to single fermion ones hence the states of such systems are
not correlated. (Alternatively, we can consider the physically
relevant interpretation of a particle-hole transformation as a
manifestation of this duality\cite{Annals}.) Similarly
$\bigwedge^3{\bf C}^5\simeq \bigwedge^2{\bf C}^5$ hence a five
dimensional three-fermion state can be mapped to a two fermion
state. Since the rank of the coefficient matrix $P_{ab}$ is always
even this case can be related to the four dimensional
one\cite{Annals} and the measure $\eta$ again can be used. Hence
as far as multipartite correlations are concerned these cases are
not interesting. The first nontrivial case is a three-fermion
system with six single particle states. From the mathematical
point of view these states can be represented by elements of
$\bigwedge^3{\bf C}^6$ the three-fold antisymmetric tensor product
of the six dimensional state space ${\bf C}^6$.

In this paper we classify different entanglement types of three
fermions with six single particle states under the SLOCC group
$GL(6, {\bf C})$. In Section II. we introduce a new tripartite
entanglement measure ${\cal T}_{123}$ quartic in the $20$
amplitudes of our fermionic state that we later show to be the
natural generalization of the well-known\cite{Kundu} three-tangle
${\tau}_{123}$ playing a central role in the classification of
three-qubit systems\cite{Dur,Miyake,Levay2}. Then in the form of
two {\it Theorems} we present our main result: two fermions with
six single particle states have four SLOCC classes, however only
two from these classes represent genuine tripartite entanglement.
The representatives of these classes will be given. Then two further
quantities of order three and two in the amplitudes are 
introduced. Taken together with ${\cal T}_{123}$ they provide
the sufficient and necessary set of quantities to determine which
class a given state belongs. We illustrate the use of these
quantities by a very simple example: two states having the same
single particle reduced density matrices, however from the
tripartite perspective they are entangled differently. Our
classification has a striking similarity to the well-known SLOCC
classification of three-qubits. In Section III. we elaborate this
point and show that this similarity is not a coincidence. We
introduce a three-qubit-like state which is a tripartite fermionic
one with only eight nonvanishing amplitudes. It is shown that our
new measure ${\cal T}_{123}$ reduces in this case to the
three-tangle ${\tau}_{123}$ based on Cayley's hyperdeterminant.
This analogy with three-qubit states enables a different
construction of our basic quantities of order four, three, and two
related to the ranks of the states appearing in the canonical
forms.

We left the proof of our {\it Theorems} to Section IV. The reason
for this is that these proofs are essentially available in the
mathematical literature, however in the somewhat exotic field of
Freudenthal triple systems and cubic Jordan algebras. The key
observation is that there is an isomorphism between these
Freudenthal triples of a special kind and our three-fermion
systems. Moreover, this isomorphism lifts equivariantly to an
isomorphism  between the invariance group of such triples and the
$SL(6, {\bf C})$ subgroup of the SLOCC group. After this
observation the SLOCC classification follows immediately from the
corresponding classification of the canonical forms of the
relevant Freudenthal triples. For convenience we also presented in
Section IV. a brief summary of the relevant concepts of these
mathematical structures. The reader interested in the details
might consult the references given in this section. In Section V.
we present our conclusions. Here we also would like to propose the
use of Pl\"ucker relations as a sufficient and necessary condition
of separability for fermionic systems with an {\it arbitrary} number
of constituents and single particle states. Using some recent
mathematical results we emphasize the basic importance of the case
of two fermions with four single particle states and the
associated three-term Pl\"ucker relation Eq. (\ref{plucky}). In
some sense the problem of separability for {\it any} fermionic
system is encoded into an equivalent two fermion system with four
single particle states. Hence as a test for separability the
measure ${\eta}$ of Eq. (\ref{eta}) is universal. With some
further comments on interesting open problems we conclude.

%**********************************************************************************************************
\section{The SLOCC classification of three-fermion systems}

Let us consider three fermions with six single particle states.
The Hilbert space for one fermion is ${\bf C}^6$, hence the total
Hilbert space is ${\cal H}\equiv {\bigwedge}^3{\bf C}^6$ i.e. the
three-fold antisymmetric tensor product of three copies of ${\bf
C}^6$. Let us introduce the notation $e^a\wedge e^b\wedge e^c$ for
the normalized Slater determinant formed from the basis vectors
$e^a, e^b, e^c$, $a,b,c=1,\dots 6$ i.e. we have
\begin{equation}
e^a\wedge e^b\wedge e^c\equiv\frac{1}{\sqrt{6}}(e^a\otimes
e^b\otimes e^c+e^c\otimes e^a\otimes e^b+e^b\otimes e^c\otimes
e^a-e^c\otimes e^b\otimes e^a-e^a\otimes e^c\otimes e^b-e^b\otimes
e^a\otimes e^c). \label{Slater}
\end{equation}
\noindent We represent a three-fermion state $\vert P\rangle$ with
six single particle states by a three-form $P\in {\bigwedge}^3{\bf
C}^3$ as
\begin{equation}
P=\frac{1}{6}\sum_{a,b,c=1}^6P_{abc}e^a\wedge e^b\wedge e^c,
\label{threeform}
\end{equation}
\noindent
 where the coefficient tensor $P_{abc}$ is totally
antisymmetric hence has $20$ independent complex components.  The
condition of normalization yields the further constraint
\begin{equation}
\vert P_{123}\vert^2+\dots+\vert P_{456}\vert^2=1. \label{norma1}
\end{equation}
\noindent Alternatively our fermionic state $\vert P\rangle$ can
be written as\cite{Schlie}

\begin{equation}
\vert P\rangle=
\sum_{a,b,c=1}^6w_{abc}f^{\dagger}_af^{\dagger}_bf^{\dagger}_c\vert
0\rangle
\label{Schstate}
\end{equation}
\noindent where $f_a$ and $f^{\dagger}_a$ are fermionic creation
and annihilation operators satisfying the usual anticommutation
relations. In this case we have $\sum_{abc}\vert
w_{abc}\vert^2=1/6$, hence $w_{abc}\leftrightarrow
P_{abc}/\sqrt{6}$.

The group of stochastic local operations and classical
communication\cite{Dur} (SLOCC) is
acting as
\begin{equation}
\vert P\rangle \mapsto (G\otimes G\otimes G)\vert P\rangle,\qquad
G\in GL(6, {\bf C}), \label{SLOCC}
\end{equation}
\noindent i.e. we are acting with the {\it same} $6\times 6$
complex invertible matrix on our three copies of ${\bf C}^6$. This
means that for the totally antisymmetric tensor $P_{abc}$ we have
\begin{equation}
P_{a_1b_1c_1}\mapsto
{G_{a_1}}^{a_2}{G_{b_1}}^{b_2}{G_{c_1}}^{c_2}P_{a_2b_2c_2}.
\label{trans}
\end{equation}
 \noindent
 We are interested in finding all the SLOCC equivalence
classes of three-fermion states. Two states are SLOCC equivalent
(hence belonging to the same class) iff their amplitudes $P_{abc}$
satisfy (\ref{trans}) for some $G\in GL(6,{\bf C})$. It is
convenient to work with the subgroup of transformations that have
unit determinant. These special SLOCC transformations are elements
of the group $SL(6, {\bf C})\subset GL(6, {\bf C})$. We first
determine the equivalence classes under $SL(6,{\bf C})$ and then
find easily how our results modify for the full group $GL(6,{\bf
C})$.

Our classification scheme is based on a new tripartite
entanglement measure invariant under the action of $G\in SL(6,
{\bf C})$ defined in Eq. (\ref{trans}).
 In order to define this measure
we reorganize the $20$ independent complex amplitudes $P_{abc}$ into two
complex numbers $\alpha,\beta$ and two complex $3\times 3$
matrices $A$ and $B$ as follows. As a first step we change our
labelling convention by using the symbols
$\overline{1},\overline{2},\overline{3}$ instead of $4,5,6$
respectively. The meaning of the labels $1,2,3$ is not changed.
Hence for example we can alternatively refer to $P_{456}$ as
$P_{\overline{1}\overline{2}\overline{3}}$ or to $P_{125}$ as
$P_{12\overline{2}}$. Now we define
\begin{equation}
\alpha\equiv P_{123},\qquad \beta\equiv P_{\overline{123}}
\label{alfabeta}
\end{equation}
\begin{equation}
A=\begin{pmatrix}A_{11}&A_{12}&A_{13}\\A_{21}&A_{22}&A_{23}\\A_{31}&A_{32}&A_{33}\end{pmatrix}
\equiv\begin{pmatrix}P_{1\overline{23}}&P_{1\overline{31}}&P_{1\overline{12}}\\
P_{2\overline{23}}&P_{2\overline{31}}&P_{2\overline{12}}\\P_{3\overline{23}}&P_{3\overline{31}}&P_{3\overline{12}}\end{pmatrix},
\label{Amatr}
\end{equation}
\begin{equation}
B=\begin{pmatrix}B_{11}&B_{12}&B_{13}\\B_{21}&B_{22}&B_{23}\\B_{31}&B_{32}&B_{33}\end{pmatrix}\equiv
\begin{pmatrix}P_{\overline{1}23}&P_{\overline{1}31}&P_{\overline{1}12}\\P_{\overline{2}23}&P_{\overline{2}31}&P_{\overline{2}12}\\
P_{\overline{3}23}&P_{\overline{3}31}&P_{\overline{3}12}\end{pmatrix}.
\label{Bmatr}
\end{equation}
Mnemonic: the row index of $A$ transforms to the first, the column
index transforms to the second and third index of $P_{abc}$ with
the {\it overlined pairs} are coming from the corresponding
complements of the column index in cyclic order. E.g. for $A_{12}$
the column label $2$ is replaced by the pair
$\overline{3}\overline{1}$. For matrix $B$ we have to form the
complement of $A$, i.e. replacing the numbers with their overlined
versions.

The new tripartite entanglement measure for fermionic systems with
six single particle states we wish to propose is
\begin{equation}
0\leq {\cal T}_{123}=\vert T_{123}\vert\leq 1 \label{ftangle}
\end{equation}
\noindent where
\begin{equation}
 T_{123}=4\left([{\rm
Tr}(AB)-\alpha\beta]^2-4{\rm Tr}(A^{\sharp}B^{\sharp})+4\alpha{\rm
Det}(A)+4\beta{\rm Det}(B)\right), \label{T}
\end{equation}
\noindent where $A^{\sharp}$ and $B^{\sharp}$ correspond to the
regular adjoint matrices for $A$ and $B$ i.e.
\begin{equation}
AA^{\sharp}=A^{\sharp}A={\rm Det }(A)I,\qquad
BB^{\sharp}=B^{\sharp}B={\rm Det }(B)I \label{sharp},
\end{equation}
\noindent with $I$ the $3\times 3$ identity matrix. The overall
factor $4$ in Eq. (\ref{T}) is chosen to ensure
 ${\cal T}_{123}\leq 1$ for normalized states.

For the classification of three-fermion states we need to define
$\vert \tilde{P}\rangle$ the {\it dual} of our three-fermion state
$\vert P\rangle$. For this we define a new tensor
$\tilde{P}_{abc}$ by defining the corresponding quantities
$(\tilde{\alpha},\tilde{\beta},\tilde{A},\tilde{B})$ via Eqs.
(\ref{alfabeta}-\ref{Bmatr})
\begin{equation}
\tilde{\alpha}=-{\alpha}^2{\beta}+\alpha{\rm Tr}(AB)-2{\rm
Det}(B), \qquad \tilde{\beta}={\alpha}{\beta}^2-\beta{\rm
Tr}(AB)+2{\rm Det}(A) \label{abedual}
\end{equation}
\begin{equation}
\tilde{A}=2B\times A^{\sharp}-2\beta B^{\sharp}-[{\rm
Tr}(AB)-\alpha\beta]A,\qquad \tilde{B}=-2A\times
B^{\sharp}+2\alpha A^{\sharp}+[{\rm Tr}(AB)-\alpha\beta]B.
\label{ABdual}
\end{equation}
\noindent
 Here for two $3\times 3$ matrices we have
\begin{equation}
M\times N\equiv(M+N)^{\sharp}-M^{\sharp}-N^{\sharp}. \label{cross}
\end{equation}
\noindent Notice that the "state"
\begin{equation}
\tilde{P}=\frac{1}{6}\sum_{a,b,c=1}^6\tilde{P}_{abc}e^a\wedge e^b\wedge e^c
\label{dualplucker}
\end{equation}
 is cubic in the original
amplitudes and it does not have to be normalized. It will turn out
that $\tilde{P}_{abc}$ has the same transformation properties as
$P_{abc}$ described by Eq. (\ref{trans}). Its role will be
clarified later.

 Now we can state the main result of
this paper.

 {\it Theorem 1:} A three-fermion state with six single
particle states can be entangled in two inequivalent ways. The two
classes with genuine tripartite entanglement are characterized by
$T_{123}\neq 0$, and $T_{123}=0$, $\tilde{P}\neq 0$. States with
$T_{123}=0$, $\tilde{P}=0$ are either separable or biseparable.

It should be clear that $T_{123}$ is a complex number but
$\tilde{P}$ is a collection of $20$ complex numbers. The condition
$\tilde{P}=0$ means that these complex numbers are all zero (i.e.
the dual state is just the zero state).

Before making a list of the representatives of each class let us
consider an example. Let us consider the following two
states\cite{Gottlieb}

\begin{equation}
\Psi =\frac{1}{\sqrt{3}}(\sqrt{2}e^1\wedge e^3\wedge e^5+e^2\wedge
e^4\wedge e^6),\qquad \Phi=\frac{1}{\sqrt{3}}(e^1\wedge e^2\wedge
e^3+e^3\wedge e^4\wedge e^5 +e^1\wedge e^5\wedge e^6).
\label{Gottlieb}
\end{equation}
It can be easily shown that the single particle reduced density
matrices corresponding to these two states are the same
\begin{equation}
\varrho_1(\Psi)=\varrho_1(\Phi)=\frac{1}{3}\begin{pmatrix}2&0&0&0&0&0\\
0&1&0&0&0&0\\0&0&2&0&0&0\\0&0&0&1&0&0\\
0&0&0&0&2&0\\0&0&0&0&0&1\end{pmatrix}
\end{equation}
hence according to the usual tests of entanglement\cite{remark}
based on a calculation of the eigenvalues of the single particle
density matrix, $\Psi$ and $\Phi$ have the same amount of
entanglement. However, for $\Psi$ we have
\begin{equation}
P_{135}=P_{13\overline{2}}=\sqrt{\frac{2}{3}},\qquad
P_{246}=P_{2\overline{1}\overline{3}}=\sqrt{\frac{1}{3}},
\end{equation}
\noindent and for $\Phi$ the corresponding quantities are
\begin{equation}
P_{123}=\frac{1}{\sqrt{3}},\quad
P_{345}=P_{3\overline{1}\overline{2}}=\frac{1}{\sqrt{3}},\quad
P_{156}=P_{1\overline{2}\overline{3}}=\frac{1}{\sqrt{3}},
\end{equation}
\noindent hence a calculation of $T_{123}$ of Eq. (\ref{T}) using Eqs.
(\ref{alfabeta}-\ref{Bmatr}) shows that
\begin{equation}
{\cal T}_{123}(\Psi)=\frac{8}{9},\qquad {\cal T}_{123}(\Phi)=0.
\label{liebtangle}
\end{equation}
\noindent Moreover, a short calculation reveals that the dual
state $\tilde{\Phi}$ is of the form
\begin{equation}
\tilde{\Phi}=\frac{2}{9}\sqrt{3} e^{\overline{2}}\wedge e^3\wedge
e^1=-\frac{2}{9}\sqrt{3}e^1\wedge e^3\wedge e^5.
\end{equation}
\noindent Since this state is not identical to zero ${\Phi}$ is
neither separable nor biseparable. Hence we conclude that the
states $\Psi$ and $\Phi$ are representatives of our two different
classes with genuine tripartite entanglement.

Now we get back to the SLOCC classification of three-fermion
states with six single particle states. We have the following

{\it Theorem 2.} Including the classes of biseparable and
separable states we have four disjoint SLOCC classes. The
representatives of these classes can be brought to the following
forms

\begin{equation}
P=\frac{1}{2}(e^1\wedge e^2\wedge e^3+e^1\wedge
e^{\overline{2}}\wedge e^{\overline{3}}+e^2\wedge
e^{\overline{3}}\wedge e^{\overline{1}}+e^3\wedge
e^{\overline{1}}\wedge e^{\overline{2}}),\quad {\cal
T}_{123}(P)\neq 0 \label{1}
\end{equation}
\noindent
\begin{equation}
P=\frac{1}{\sqrt{3}}(e^1\wedge e^2\wedge e^3+e^1\wedge
e^{\overline{2}}\wedge e^{\overline{3}}+e^2\wedge
e^{\overline{3}}\wedge e^{\overline{1}}),\qquad {\cal
T}_{123}(P)=0,\quad \tilde P\neq 0
\end{equation}
\noindent
\begin{equation}
P=\frac{1}{\sqrt{2}}(e^1\wedge e^2\wedge e^3+e^1\wedge
e^{\overline{2}}\wedge e^{\overline{3}}),\qquad {\cal
T}_{123}(P)=0,\quad \tilde{P}=0
\end{equation}
\noindent
\begin{equation}
P=e^1\wedge e^2\wedge e^3,\qquad {\cal T}_{123}(P)=0,\quad
\tilde{P}=0. \label{4}
\end{equation}
\noindent Obviously the last two classes correspond to biseparable
and separable states. The representative of the second class is
very similar to the state $\Phi$ of Eq. (\ref{Gottlieb}). For the
representative of the first class we prefer the four term form,
but we will show later that this class can alternatively be
represented by a two-term expression (as we also expect from our
study with the state $\Psi$ of Eq. (\ref{Gottlieb})). Notice also
the striking similartity with the well-known SLOCC classification
obtained for three-qubit states\cite{Dur}. This is not a
coincidence as we will show in the next section.

In order to complete our classification we should find a means of
deciding whether a state having {\it no} genuine tripartite
entanglement is separable or biseparable. Indeed, the conditions
${\cal T}_{123}(P)=0$ and $\tilde{P}=0$ do not specify whether our
state is totally separable or merely biseparable.

Let us call a fermionic state $\vert P\rangle$ {\it separable} if
the corresponding form $P\in {\bigwedge}^3{\bf C}^6$ is {\it
decomposable} i.e. if it can be written as
$P={\omega}_1\wedge{\omega}_2\wedge{\omega}_3 $ for some
${\omega}_j\in {\bf C}^6, j=1,2,3$. As it is well-known in
multilinear algebra\cite{Harris,Hodge} the form $P$ is
decomposable if and only if the
\begin{equation}
{\Pi}_{{\cal A},{\cal B}}(P)\equiv
\sum_{i=1}^4(-1)^{i-1}P_{a_1a_2b_i}P_{b_1b_2b_3b_4{\hat{b}}_i}=0
\label{pluckerrel}
\end{equation}
\noindent {\it Pl\"ucker relations} hold, for all ${\cal
A}\equiv\{a_1,a_2\}$ two and ${\cal B}\equiv\{b_1,b_2,b_3,b_4\}$
four element subsets of the set $\{1,2,3,4,5,6\}$ (or
alternatively the one
$\{1,2,3,\overline{1},\overline{2},\overline{3}\}$). Here
${\hat{b}}_i$ means that we have to delete  $b_i$ from the list
$b_1,b_2,b_3,b_4$. The Pl\"ucker relations define a set of
quadratic forms labelled by all possible subsets ${\cal A}$ and
${\cal B}$ compatible with the antisymmetry properties of the $20$
components $P_{abc}$. An excercise in combinatorics\cite{Kasman}
shows that the number of possible quadratic forms is $45$.

As an example let us consider the state
\begin{equation}
\Omega=\frac{1}{2}(e^1\wedge e^4\wedge e^5-e^1\wedge e^4\wedge e^6
+e^3\wedge e^4\wedge e^5-e^3\wedge e^4\wedge e^6),
\end{equation}
\noindent with $P_{145}=P_{345}=-P_{146}=-P_{346}=1/2$. The only
subset combinations to be checked are $(\{14\},\{3456\})$,
$(\{34\},\{1456\})$, $(\{45\}.\{1346\})$, and $(\{46\},\{1345\})$,
all of them give the same associated quadratic form, which is
zero. Hence this state is separable, as it has to be since $\Omega
=\frac{1}{2}(e^1+e^3)\wedge e^4\wedge (e^5-e^6)$.

Now we can complete our classification scheme by identifying the
{\it biseparable} states as the ones having ${\cal
T}_{123}(P)=0,\tilde{P}=0,\Pi(P)\neq 0$, and the {\it separable}
ones with ${\cal T}_{123}(P)=0,\tilde{P}=0,\Pi(P)=0$. Here
$\Pi(P)=0$ refers to the vanishing of the relevant $45$ quadratic
combinations of the $20$ independent amplitudes $P_{abc}$.

\section{An analogy with three qubit systems}

In the previous section we have found a striking similarity
between our SLOCC classification of three-fermion states and the
one for three-qubit states\cite{Dur}. Now, by studying a special
subset of three-fermion systems we show that our classification
can indeed be regarded as a generalization of the well-known
results obtained for three qubits. The subsystem we wish to study
has merely $8$ complex amplitudes. In the notation of Eqs.
(\ref{alfabeta}-\ref{Bmatr}) these nonvanishing amplitudes are
arranged as
\begin{equation}
\alpha=P_{123},\quad\beta=P_{\overline{123}},\quad
A=\begin{pmatrix}P_{1\overline{23}}&0&\\0&P_{2\overline{31}}&0\\0&0&P_{3\overline{12}}\end{pmatrix}
,\quad
B=\begin{pmatrix}P_{\overline{1}23}&0&0\\0&P_{\overline{2}31}&0\\0&0&P_{\overline{3}12}\end{pmatrix}.
\label{3qubit}
\end{equation}
\noindent Due to the antisymmetry properties of the tensor
$P_{abc}$ we can arrange all these amplitudes to have the $1$ or
$\overline{1}$ in the first, the $2$ and $\overline{2}$ in the
second, and the $3$ and $\overline{3}$ in third position. Hence we
have a state with a collection of $8$ complex amplitudes
$(P_{123},P_{12\overline{3}},P_{1\overline{2}3},P_{1\overline{23}},
P_{\overline{1}23},P_{\overline{1}2\overline{3}},P_{\overline{12}3},P_{\overline{123}})$.
Let us denote this new state by $\vert{\cal P}\rangle$ and the
associated $3$-form by
\begin{equation}
{\cal P}=P_{123}e^1\wedge e^2\wedge
e^3+P_{12\overline{3}}e^1\wedge e^2\wedge e^{\overline{3}}+\dots
+P_{\overline{123}}e^{\overline{1}}\wedge e^{\overline{2}}\wedge
e^{\overline{3}}. \label{3qubitform}
\end{equation}
\noindent Let us compare this with the usual expression for a
three-qubit state
\begin{equation}
\psi=\psi_{000}e^0\otimes e^0\otimes e^0+\psi_{001}e^0\otimes
e^0\otimes e^1+\dots +\psi_{111}e^1\otimes e^1\otimes e^1
\end{equation}
\noindent or in the usual notation of quantum information theory
\begin{equation}
\vert\psi\rangle =\psi_{000}\vert 000\rangle +\psi_{001}\vert
001\rangle +\dots +\psi_{111}\vert 111\rangle. \label{usual3qubit}
\end{equation}
\noindent  We see that if the indices $1,2,3$ refer to subsystem
labels of some fictious system and the lack of overbar corresponds
to $0$ and an overbar corresponds to $1$ we have a mapping between
the three-qubit states and our {\it special} three fermion states.

Let us work out how $T_{123}(\cal P)$ looks like. In order to make
expressions more transparent by an abuse of notation we apply the
instructive three-qubit labelling, hence we have $P_{123}\equiv
P_{000}$, $P_{12\overline{3}}\equiv P_{001}$ e.t.c. A further
simplification can be obtained by reverting to decimal notation,
i.e. $P_{123}\equiv P_0$, $P_{12\overline{3}}\equiv P_1,\dots
,P_{\overline{123}}\equiv P_7$. Using this notation the final
expression for $T_{123}({\cal P})$ is
\begin{equation}
T_{123}=4D({\cal P})
\end{equation}
\noindent where
\begin{eqnarray}
D({\cal P})&=&(P_0P_7)^2+(P_1P_6)^2+(P_2P_5)^2+(P_3P_4)^2
-2(P_0P_7)[(P_1P_6)+(P_2P_5)+(P_3P_4)]\nonumber\\
&-&2[(P_1P_6)(P_2P_5)+(P_2P_5)(P_3P_4)+(P_3P_4)(P_1P_6)]\nonumber\\
&+&4P_0P_3P_5P_6+4P_7P_4P_2P_1
\end{eqnarray}
\noindent
is Cayley's hyperdeterminant\cite{Cayley,Gelfand}. It is
related to the {\it three-tangle}\cite{Kundu} ${\tau}_{123}$ the
canonical measure of tripartite entanglement as $\tau_{123}=4\vert
D({\cal P})\vert$. Hence for a normalized special three-fermion
state we have
\begin{equation}
0\leq {\cal T}_{123}({\cal P})={\tau}_{123}({\cal P})\leq 1.
\label{reduction}
\end{equation}

Now it is easy to understand the similarity between our
classification as presented by {\it Theorem 2.} and the usual one
for three-qubit systems. Mapping the representative states of Eqs.
(\ref{1}-\ref{4}) of {\it Theorem 2.} to a corresponding three
qubit one we get the four possibilities
\begin{equation}
\frac{1}{2}(\vert 000\rangle+\vert 011\rangle +\vert 101\rangle
+\vert 110\rangle) \label{1v}
\end{equation}
\begin{equation}
\frac{1}{\sqrt{3}}(\vert 000\rangle +\vert 011\rangle +\vert
101\rangle)
\end{equation}
\begin{equation}
\frac{1}{2}(\vert 000\rangle +\vert 011\rangle)
\end{equation}
\begin{equation}
\vert 000\rangle. \label{4v}
\end{equation}
\noindent It is easy to show that
\begin{equation}
\frac{1}{2}(\vert 000\rangle +\vert 011\rangle +\vert 101\rangle
+\vert 110\rangle)=(H\otimes H\otimes H)\frac{1}{\sqrt{2}}(\vert
000\rangle +\vert 111\rangle)=H\otimes H\otimes H\vert GHZ\rangle,
\label{GHZ}
\end{equation}
\noindent and
\begin{equation}
\frac{1}{\sqrt{3}}(\vert 000\rangle +\vert 011\rangle +\vert
101\rangle)=(I\otimes I\otimes X)\frac{1}{\sqrt{3}}(\vert
001\rangle+\vert 010\rangle +\vert 100\rangle)=(I\otimes I\otimes
X)\vert W\rangle. \label{W}
\end{equation}
\noindent
 where $H$ and $X$ are  the usual
Hadamard and bit flip gates
\begin{equation}
H=\frac{1}{\sqrt{2}}\begin{pmatrix}1&1\\1&-1\end{pmatrix},\qquad
X=\begin{pmatrix}0&1\\1&0\end{pmatrix}. \label{gates}
\end{equation}
Hence, these states are local unitary (hence also SLOCC)
equivalent to the usual $GHZ$ and $W$ states\cite{Dur}. Notice
also that since\cite{Gelfand}
\begin{equation}
D((G_1\otimes G_2\otimes G_3)\psi)={\rm Det}(G_1)^2{\rm
Det}(G_2)^2{\rm Det}(G_3)^2D(\psi),\qquad G_1,G_2,G_3\in GL(2,{\bf
C})\label{trafocayley}
\end{equation}
\noindent none of these transformations changes the value of
Cayley's hyperdeterminant.

Now in order to complete our demonstration that the three-qubit
SLOCC classification is naturally embedded into the one of {\it
Theorem 2.} we have to also show that the SLOCC group $GL(2, {\bf
C})^{\otimes 3}$ is indeed embedded into our $GL(6, {\bf C})$. It
is obvious that this embedding is effected by looking at that
subgroup of $GL(6, {\bf C})$ that leaves invariant the special
form of the state ${\cal P}$ of Eq. (\ref{3qubitform}). Such
states are clearly the ones leaving the subspaces $1\overline{1}$,
$2\overline{2}$ and $3\overline{3}$ invariant. The reduction of an
element of $GL(6, {\bf C})$ hence contains three $2\times 2$
blocks of $GL(2, {\bf C})$ transformations $G_1,G_2$ and $G_3$. By
virtue of the antisymmetry property of the tensor $P_{abc}$ it is
now easy to see that the action of Eq. (\ref{trans}) gives rise to
the usual one of the form $G_1\otimes G_2\otimes G_3$.

This embedding of the three-qubit system into our three-fermion
one is also useful to find an alternative expression for our new
tripartite entanglement measure ${\cal T}_{123}$. First recall the
alternative expression\cite{Kundu} for Cayley's hyperdeterminant
\begin{equation}
D(\psi)=-\frac{1}{2}{\epsilon}^{A_1A_3}{\epsilon}^{A_2A_4}{\epsilon}^{B_1B_2}
{\epsilon}^{C_1C_2}{\epsilon}^{B_3B_4}{\epsilon}^{C_3C_4}\psi_{A_1B_1C_1}
\psi_{A_2B_2C_2}\psi_{A_3B_3C_3}\psi_{A_4B_4C_4}
 \label{Cayleyeps}
\end{equation}
\noindent where $A_1,\dots C_3=0,1$. Proceeding by analogy the
relevant expression we have found for $T_{123}$ is
\begin{equation}
T_{123}=-\frac{1}{6^3}{\epsilon}^{a_1b_1c_1a_3b_2c_2}{\epsilon}^{a_2b_3c_3a_4b_4c_4}
P_{a_1b_1c_1}P_{a_2b_2c_2}P_{a_3b_3c_3}P_{a_4b_4c_4},
\label{newtangle}
\end{equation}
\noindent where now $a_1,\dots c_4=1,2,\dots 6$. Notice that this
expression can be written in the form
\begin{equation}
 T_{123}=-\frac{1}{3}{\epsilon}^{abcdef}P_{abc}\tilde{P}_{def},
\label{szepalak}
\end{equation}
\noindent where $\tilde{P}$ is the dual state introduced in
Eq.(\ref{abedual})-(\ref{ABdual}), with its new form
\begin{equation}
\tilde{P}_{abc}=\frac{1}{72}{\epsilon}^{klmk^{\prime}l^{\prime}m^{\prime}}P_{alm}P_{kbc}P_{k^{\prime}l^{\prime}m^{\prime}}.
\label{dualeps}
\end{equation}
\noindent Notice that Eq.(\ref{szepalak}) can be used to define
the antisymmetric (symplectic) form
\begin{equation}
\{\cdot,\cdot\}:(P,Q)\mapsto
\{P,Q\}\equiv\frac{1}{6}{\varepsilon}^{abcdef}P_{abc}Q_{def}\in
{\bf C}. \label{simplectic}
\end{equation}
\noindent It should be clear that if $P_{abc}$ and $Q_{abc}$ are
transforming according to Eq. (\ref{trans}) the symplectic form is
invariant under the $SL(6, {\bf C})$ subgroup of the SLOCC group,
and transforms by picking up a factor proportional to the
determinant for the full SLOCC group.

Now we can neatly summarize the quantities and the role they are
playing in our SLOCC classification. We need three quantities, of
order four, three and two in the amplitudes $P_{abc}$. They are
given by Eqs. (\ref{newtangle}), (\ref{dualeps}) and
(\ref{pluckerrel}). The four different SLOCC canonical forms have
four, three, two and one terms (see Eqs. (\ref{1}-\ref{4})). We
call these classes of {\it rank} four, three, two and one
respectively. The rank equals four iff $T_{123}\neq 0$. The rank
is less than or equal to three iff $T_{123}=0$, less than or equal
to two iff $\tilde{P}=0$. Finally the Pl\"ucker relation gives
the result that the rank is less than or equal to one iff
${\Pi}(P)=0$.

\section{Cubic Jordan algebras and Freudenthal triples}

As we have already discussed in the Introduction the proof of both
of our theorems is available in the mathematics literature.
However, these results are scattered in the exotic domain of
mathematics of Freudenthal triple systems and cubic Jordan
algebras, concepts that have not made their debut to quantum
information theory yet. The only notable exception where these
algebraic structures play some role is the current research topic
called "black hole analogy" where mathematical connections between
stringy black hole solutions and the theory of quantum
entanglement have been established\cite{Levay,Duff1}. Luckily in
order to understand the basic correspondence between such
algebraic constructs and our fermionic systems one does not have
to dwell deep into the subject. Here we merely streamline the
basic ideas of the proof, the interested reader should consult the
literature\cite{Mc}.

A Jordan algebra ${\cal J}$ over a field ${\bf F}$ (we have the
complex numbers in our mind) is a vector space $V$ over ${\bf F}$
with a bilinear product $\circ$ ({\it Jordan product}) satisfying
the axioms
\begin{equation}
A\circ B=B\circ A,\qquad A^2\circ(A\circ B)=A\circ(A^2\circ B)
,\qquad A,B\in{\cal J}, \label{jaxiom}
\end{equation}
\noindent here $A^2\equiv A\circ A$. The product is commutative by
definition but it does not have to satisfy the associative law.
The only example that we need in this paper is the trivial one of
 the Jordan algebra of $3\times 3$ matrices  with complex elements
denoted by $M(3,{\bf C})$. The Jordan product in this case is
defined as
\begin{equation}
A\circ B\equiv\frac{1}{2}(AB+BA),\qquad A,B\in M(3,{\bf C}).
\label{product}
\end{equation}
Here $AB$ refers to the usual (associative) product of the
relevant matrices. We will be interested in the so called {\it
cubic Jordan algebras}, the ones in which every element satisfies
a cubic polynomial equation. In our case $M(3, {\bf C})$ is
obviously a cubic Jordan algebra since by Cayley-Hamilton we have
\begin{equation}
A^3-{\rm Tr}(A)A^2+\frac{1}{2}({\rm Tr}(A)^2-{\rm Tr}(A^2))A-{\rm
Det}(A)I=0. \label{cubic}
\end{equation}
\noindent In $M(3,{\bf C})$ regarded as a Jordan algebra ${\rm
Det}(A)$ is called the (cubic) {\it norm} of $A$ denoted also by
$N(A)$. Moreover, a bilinear form $(\cdot,\cdot):M(3,{\bf
C})\times M(3,{\bf C})\to {\bf C}$ can also be defined by
\begin{equation}
(A,B)={\rm Tr}(A\circ B)={\rm Tr}(AB). \label{bili}
\end{equation}
\noindent
 One can
uniquely define the quadratic {\it adjoint/sharp } map by
\begin{equation}
(A^{\sharp},B)=3N(A,A,B), \label{sharp2}
\end{equation}
using the linearization of the norm\cite{Mc}. For our case it
turns out that $A^{\sharp}$ is just the usual adjoint (transposed
cofactor) matrix familiar from Eq. (\ref{sharp}), and the map of
Eq. (\ref{cross}) is just the linearization of the sharp map.

Note that in the general case we can construct cubic Jordan
algebras via the so called Springer construction\cite{Mc}. In this
case one starts with a vector space $V$ with a cubic form $N:V\to
{\bf F}$ and a special point $c$ called the base point (in our
special case it is just the identity matrix $I$). Then via
linearization of $N$ one defines suitable linear, quadratic,
bilinear  and trace bilinear maps that give rise to the definition
of the sharp map and its linearization. If the trace bilinear form
is nondegenerate, and the sharp map satisfies
$(A^{\sharp})^{\sharp}=N(A)A$ then we have a {\it Jordan cubic}.
Then it is proved that every Jordan cubic gives rise to a Jordan
algebra with unit $1\equiv c$. The Jordan product is given by an
explicit formula in terms of the linearization of the sharp map,
the linear and the bilinear maps\cite{Mc}. In the following we do
not need this general construction of cubic Jordan algebras
however, it is important to bear in mind that the constructions
yielding the canonical forms we are going to describe are valid
even in this general case.

Next we define the {\it structure group} of the cubic Jordan
algebra ${\cal J}$ as the set of invertible ${\bf F}$ linear
transformations $g$ of the vector space ${\cal J}$ which preserve
the norm up to a scalar $\lambda\in {\bf F}$ which depends on $g$
only,
\begin{equation}
{\rm Str}({\cal J})=\{g\in GL({\cal J})\vert N(gA)=\lambda(g)N(A),
 A\in {\cal J}\}. \label{str}
\end{equation}
\noindent \noindent For $M(3, {\bf C})$ the structure group ${\rm
Str}({\cal J})$ is generated by transformations of the form
\begin{equation}
h:A\mapsto \Lambda_1A\Lambda_2^{-1},\qquad \Lambda_1,\Lambda_2\in
GL(3,{\bf C}),\quad A\in M(3,{\bf C}), \label{str2}
\end{equation}
\noindent and $t: A\mapsto A^T$ where $A^T$ refers to the
transpose of $A$. We denote by ${\rm Str}_{\circ}({\cal J})$ the
component connected to the identity of ${\rm Str}({\cal J})$
generated by the transformations $h$ of Eq. (\ref{str2}).

Now we can define a Freudenthal triple
system\cite{Brown,Faulkner,Freudenthal}. This is a vector space
${\cal M}={\cal M}({\cal J})$ constructed from the cubic Jordan
algebra in the following way
\begin{equation}
{\cal M}({\cal J})={\bf F}\oplus{\bf F}\oplus {\cal J}\oplus{\cal
J}. \label{freund}
\end{equation}
\noindent Obviously ${\rm dim}{\cal M}=2+2{\rm dim}{\cal J}$. In
our case we have
\begin{equation}
{\cal M}={\bf C}\oplus{\bf C}\oplus M(3,{\bf C})\oplus M(3, {\bf
C}), \label{freudkomplex}
\end{equation}
\noindent with complex dimension $2\times 9+2=20$. An element of
${\cal M}$ can be written as
\begin{equation}
x=\begin{pmatrix}\alpha&A\\B&\beta\end{pmatrix},\qquad
\alpha,\beta\in {\bf C},\quad A,B\in M(3,{\bf C}). \label{x}
\end{equation}
\noindent Notice that the quantity $x$ can alternatively be used
as a shorthand notation for our quantities introduced in Eqs.
(\ref{alfabeta}-\ref{Bmatr}) related to the $20$ amplitudes of our
fermionic states.

On ${\cal M}$ there are two important extra structures: a
skew-symmetric (symplectic) bilinear form $\{\cdot,\cdot\}:{\cal
M}\times {\cal M}\to {\bf F}$, and a quartic form $q:{\cal M}\to
{\bf M}$ defined by
\begin{equation}
\{x,y\}=\alpha\delta-\beta\gamma+(A,D)-(B,C),\quad
x=\begin{pmatrix}\alpha &A\\B &\beta\end{pmatrix},\quad
y=\begin{pmatrix}\gamma&C\\D&\delta\end{pmatrix}, \label{sympl}
\end{equation}
\noindent
\begin{equation}
q(x)=2\left((A,B)-\alpha\beta\right)^2-8(A^{\sharp},B^{\sharp})+8\alpha
N(A)+8\beta N(B). \label{q}
\end{equation}
\noindent After recalling that for our cubic Jordan algebra $M(3,
{\bf C})$ we have $N(A)={\rm Det}(A)$ and $(A,B)={\rm Tr}(AB)$ we
see that the tripartite entanglement measure of Eqs.
(\ref{ftangle}-\ref{T}) is related to this quartic form as
\begin{equation}
{\cal T}_{123}=2\vert q(x)\vert, \label{tangfreud}
\end{equation}
with $x$ given by Eq.(\ref{x}) and Eqs.
(\ref{alfabeta}-\ref{Bmatr}). Moreover, if we associate to the
pair $(x,y)$ occurring in Eq.(\ref{sympl}) the one $(P,Q)$ of
three-fermion states we get
\begin{equation}
\{x,y\}=\frac{1}{6}{\varepsilon}^{abcdef}P_{abc}Q_{def},
\label{symplfermi}
\end{equation}
\noindent which is just the symplectic form of Eq.
(\ref{simplectic}). Recalling the definition of the dual
three-fermion state Eq. (\ref{dualeps}) we see that Eq.
(\ref{newtangle}) can be written as
\begin{equation}
T_{123}=-2\{x,\tilde{x}\}=2\{\tilde{x},x\}=2q(x),\qquad
\tilde{x}=\begin{pmatrix}\tilde{\alpha}&\tilde{A}\\
\tilde{B}&\tilde{\beta}\end{pmatrix}, \label{freualak}
\end{equation}
\noindent where for the definition of the quantities
$\tilde{\alpha},\tilde{\beta},\tilde{A},\tilde{B}$ see Eqs.
(\ref{abedual}-\ref{ABdual}). In the theory of Freudenthal triples
$\tilde{x}$ (our dual fermionic state) corresponds to the so
called trilinear map\cite{Brown} $T:{\cal M}\times {\cal M}\times
{\cal M}\to {\bf F}$ related to the quartic form as
$q(x)=\{T(x,x,x),x\}$.

The invariance group of the Freudenthal triple ${\rm Inv}({\cal
M})$ is the group of invertible ${\bf F}$ linear transformations
preserving the symplectic and quartic forms, i.e.
\begin{equation}
\{gx,gy\}=\{x,y\},\qquad q(gx)=q(x),\qquad g\in{\rm Inv}({\cal
M}),\quad x,y\in{\cal M}. \label{inv}
\end{equation}
\noindent The structure of this group has been studied for example
by Brown\cite{Brown}. It was shown that ${\rm Inv}({\cal M})$ is
generated by elements of three basic types. We give these
generators for the case interesting to us i.e. ${\cal M}({\cal
J})$ where ${\cal J}=M(3,{\bf C}).$ The component connected to the
identity ${\rm Inv}_0({\cal M})$ of ${\rm Inv}({\cal M})$ is
generated by $\sigma(\Lambda)$, $\pi(\Lambda)$ and
$\varrho(\Lambda_1,\Lambda_2)$ where $\Lambda\in M(3,{\bf C})$ and
$\Lambda_1,\Lambda_2\in GL(3,{\bf C})$. The action of these
transformations on $x\in {\cal M}$ takes the following
form\cite{Brown,Krutelevich}
\begin{equation}
\sigma(\Lambda):\begin{pmatrix}\alpha&A\\B&\beta\end{pmatrix}\mapsto
\begin{pmatrix}\alpha +(B,\Lambda)+(A,\Lambda^{\sharp})+\beta
N(\Lambda)&A+\beta\Lambda\\
B+A\times\Lambda+\beta\Lambda^{\sharp}&\beta\end{pmatrix},\quad\Lambda\in
M(3,{\bf C}) \label{sigma}
\end{equation}
\noindent
\begin{equation}
\pi(\Lambda):\begin{pmatrix}\alpha&A\\B&\beta\end{pmatrix}\mapsto\begin{pmatrix}\alpha&
A+B\times\Lambda+\alpha\Lambda^{\sharp}\\
B+\alpha\Lambda& \beta+(A,\Lambda)+(B,\Lambda^{\sharp})+\alpha
N(\Lambda)\end{pmatrix}, \qquad \label{pi}
\end{equation}
\noindent
\begin{equation}
\varrho(\Lambda_1,\Lambda_2):\begin{pmatrix}\alpha&A\\B&\beta\end{pmatrix}\mapsto
\begin{pmatrix}\frac{{\rm Det}(\Lambda_2)}{{\rm
Det}(\Lambda_1)}\alpha&\Lambda_1A\Lambda_2^{-1}\\
\Lambda_2B\Lambda_1^{-1}&\frac{{\rm Det}(\Lambda_1)}{{\rm
Det}(\Lambda_2)}\beta\end{pmatrix},\qquad \Lambda_1,\Lambda_2\in
GL(3,{\bf C}). \label{ro}
\end{equation}
\noindent The total group ${\rm Inv}({\cal M})$ is obtained by
including in $\varrho$ also the discrete transformation $x\mapsto
x^{\prime}$ by transforming merely $A$ and $B$ by taking their
transpose. (See the discussion on the structure group of ${\cal
J}$ following Eq. (\ref{str2})).

The key theorem for the SLOCC classification of our fermionic
systems has been proved in the nice paper of
Krutelevich\cite{Krutelevich}. It states that every element of
${\cal M}$ is ${\rm Inv}({\cal M})$ equivalent to one of the
following "canonical" forms
\begin{equation}
\begin{pmatrix}1&{\rm diag}\{0,0,0\}\\0&0\end{pmatrix},\quad
\begin{pmatrix}1&{\rm diag}\{1,0,0\}\\0&0\end{pmatrix},\quad
\begin{pmatrix}1&{\rm diag}\{1,1,0\}\\0&0\end{pmatrix},\quad
\begin{pmatrix}1&{\rm diag}\{1,1,k\}\\0&0\end{pmatrix}
\end{equation}
\noindent where $k\in {\bf C}$, $k\neq 0$. The four cases
correspond to the ones based on the concept of rank we
introduced at the end of Section III. Notice also that using the
correspondence given by Eqs. (\ref{alfabeta}-\ref{Bmatr}) this
classification nearly gives our classification of ${\it Theorem
2.}$ The only subtlety arising is that the rank four case gives an
infinity of subclasses labelled by the nonzero complex number $k$.
These unnormalized states are of the form
\begin{equation}
e^1\wedge e^2\wedge e^3+e^1\wedge e^{\overline{2}}\wedge
e^{\overline{3}}+e^{\overline{1}}\wedge e^2\wedge
e^{\overline{3}}+ke^{\overline{1}}\wedge e^{\overline{2}}\wedge
e^3. \label{k}
\end{equation}
\noindent However, we have not used the full SLOCC group yet. The
canonical forms obtained for ${\cal M}$ use the group ${\rm
Inv}_0({\cal M})$ which turns out to be isomorphic to $SL(6,{\bf
C})$ modulo its center\cite{Krutelevich}. Hence we still have the
freedom to rescale our states by using the full SLOCC group
$GL(6,{\bf C})$. From Eq. (\ref{T}) we see that for our state of
Eq. (\ref{k}) $T_{123}=16k$, moreover from the alternative
expression of Eq. (\ref{szepalak}) we see that $T_{123}$ picks up
a factor corresponding to the determinant of the transformation.
Hence we can use this extra freedom to achieve $T_{123}=1$ and the
canonical form of {\it Theorem 2}.

In order to make the correspondence between Freudenthal systems
based on the cubic Jordan algebra $M(3, {\bf C})$ and fermionic
systems with six single particle states precise we have to also
describe the correspondence between their relevant invariance
groups, i.e. ${\rm Inv}_0({\cal M})$ and the SLOCC subgroup $SL(6,
{\bf C})$. Let us define $\omega_6=e^{2\pi i/6}$ and
$\omega_3=e^{2\pi i/3}$ the sixth and third roots of unity. Then
$SL(6,{\bf C})$ clearly has a center $\omega_6 I_6$, where $I_6$ is
the six dimensional identity matrix. Moreover, $SL(6, {\bf C})$
transformations of the form
\begin{equation}
\vert P\rangle \mapsto
(\omega_3I_6)\otimes(\omega_3I_6)\otimes(\omega_3I_6)\vert
P\rangle
\end{equation}
\noindent leave the state $\vert P\rangle$ invariant. Hence $SL(6,
{\bf C})/\omega_3I_6$ acts on ${\cal H}=\bigwedge^3{\bf C}^6$
faithfully. Now the dictionary of Eqs.
(\ref{alfabeta}-\ref{Bmatr}) provides an isomorphism between the
$20$ complex dimensional  vector spaces ${\cal M}$ and ${\cal H}$.
Let us denote this isomorphism by $f:{\cal M}\to {\cal H}$. Now
what we need is also an associated isomorphism  $F$ of groups $F:{\rm
Inv}_0({\cal M})\to SL(6, {\bf C})/\omega_3I_6$ satisfying
\begin{equation}
f(gx)=F(g)\cdot f(x),\quad g\in {\rm Inv}_0({\cal M}),\quad x\in
{\cal M}. \label{ekvivariancia}
\end{equation}
\noindent Since ${\rm Inv}_0({\cal M})$ is generated by three
different classes of elements, we have to give the image of these
generators under $F$ satisfying Eq. (\ref{ekvivariancia}). One can
check that the relevant map is\cite{Krutelevich}
\begin{equation}
F:\sigma(\Lambda)\mapsto\begin{pmatrix}I&0\\\Lambda
&I\end{pmatrix},\qquad
F:\pi(\Lambda^{\prime})\mapsto\begin{pmatrix}I&\Lambda^{\prime}\\0&I\end{pmatrix},\qquad
\Lambda,\Lambda^{\prime}\in M(3,{\bf C}) \label{elso}
\end{equation}
\noindent
\begin{equation}
F:\varrho(\Lambda_1,\Lambda_2)\mapsto\lambda_1\lambda_2\begin{pmatrix}\Lambda_1/{\rm
Det}(\Lambda_1)&0\\0&\Lambda_2/{\rm
Det}(\Lambda_2)\end{pmatrix},\qquad \lambda^3_j={\rm
Det}(\Lambda_j), \quad  j=1,2 \label{masodik}
\end{equation}
\noindent where $\Lambda_1,\Lambda_2\in GL(6, {\bf C})$.

Let us now consider the special case when
\begin{equation}
\Lambda=\begin{pmatrix}\mu_1&0&0\\0&\mu_2&0\\0&0&\mu_3\end{pmatrix},
\qquad
\Lambda^{\prime}=\begin{pmatrix}\nu_1&0&0\\0&\nu_2&0\\0&0&\nu_3\end{pmatrix},
\qquad\Lambda_1=\Lambda_2^{-1}=\begin{pmatrix}\lambda_1&0&0\\0&\lambda_2&0\\0&0&\lambda_3\end{pmatrix}
\label{sl2c}
\end{equation}
\noindent where $\mu_1,\dots \nu_3\in {\bf C}$, and
$\lambda_1,\lambda_2,\lambda_3\in {\bf C}-\{0\}$. The
transformations associated to these parameters clearly map a
three-qubit-like state of type $\vert{\cal P}\rangle$ of Eq.
(\ref{3qubitform}) to the same type. These transformations leave
invariant the subspaces $(1\overline{1})$, $(2\overline{2})$, and
$(3\overline{3})$. The $2\times 2$ matrices operating on these
subspaces are of the form
\begin{equation}
\begin{pmatrix}1&0\\\mu_{1,2,3}&1\end{pmatrix},\qquad
\begin{pmatrix}1&\nu_{1,2,3}\\0&1\end{pmatrix},\qquad
\begin{pmatrix}\lambda_{1,2,3}&0\\0&\lambda_{1,2,3}^{-1}\end{pmatrix}.
\label{sl2c2}
\end{equation}
\noindent These generate three copies of the group $SL(2,
{\bf C})$ i.e. the SLOCC subgroup of determinant one
transformations on three qubits. Writing out explicitly the action
of this subgroup of $SL(6,{\bf C})$ on states of type $\vert{\cal
P}\rangle$ we recover the usual law
\begin{equation}
\vert{\cal P}\rangle\mapsto (S_1\otimes S_2\otimes S_3)\vert{\cal
P}\rangle,\qquad S_1,S_2,S_3\in SL(2, {\bf C}).
\label{3qubittrans}
\end{equation}
This sheds some more light on our explanation of the
three-qubit-like structure embedded in the three fermion system we
have found in Section III. (See the discussion following Eq.
(\ref{trafocayley})).

\section{Conclusions}
In this paper we investigated the entanglement properties of
three-fermion systems with six single particle states. For such
systems we introduced a new measure ${\cal T}_{123}$ of tripartite
entanglement (Eq.(\ref{ftangle}) ) depending on the $20$ complex
amplitudes characterizing our fermionic state. This entanglement
measure is of quartic order, and can be regarded as a
generalization of the well-known three-tangle ${\tau}_{123}$ of
three-qubit entanglement based on Cayley's hyperdeterminant. We
also introduced two further quantities of order three  and two in
the amplitudes (see Eqs (16-\ref{dualplucker},\ref{pluckerrel})).
They are the dual fermionic state ${\tilde{P}}$ and the Pl\"ucker
relations ${\Pi}(P)$ . Using these three quantities in concert we
managed to obtain the SLOCC classification of our three fermion
systems. We have four SLOCC classes. Apart from the separable and
biseparable ones we have two nontrivial classes with tripartite
entanglement. The canonical forms of these classes are given by
Eqs. (\ref{1}-\ref{4}). These states are the representatives of
the corresponding four classes. For the number of terms appearing
in the canonical form we coined the term {\it rank}. States with
${\cal T}_{123}\neq 0$ are of rank four, the ones with ${\cal
T}_{123}=0$ have rank at most three, the ones with
$\tilde{P}$=0 at most two, and at last fermionic states with
$\Pi(P)=0$ are of rank one. This notion of rank obviously
generalizes the concept of Slater rank well-known from the
corresponding classification of bipartite fermionic systems.

We have found a striking similarity between our SLOCC
classification and the corresponding one obtained for three-qubit
systems. This is not a coincidence. By employing a special
three-qubit-like fermionic state with $8$ amplitudes we managed to
demonstrate that the three-qubit SLOCC classification is naturally
incorporated within the fermionic one. By restriction to the state
with merely  $8$ amplitudes ${\cal T}_{123}$ reduces to
${\tau}_{123}$. This phenomenon is similar to the one found by
Gittings and Fischer\cite{Gittings} for systems of two fermions
with four single particle states. In this case it is easy to see
that the fermionic measure ${\eta}$ of Eq. (\ref{eta})  reduces to
the two-qubit {\it concurrence} ${\cal C}$. This analogy enabled
an alternative construction for our quantities of order four ,
three and two providing additional insight into their structure (Eqs. 30, 48, 50).
Finally we highlighted the proof of our theorems via introducing
the reader to the basics of cubic Jordan algebras and Freudenthal
triple systems. For the proof we referred to existing results in
the mathematical literature.

This unexpected  connection between such algebraic constructs and
quantum entanglement might prove to be useful to obtain a
classification of further special entangled systems. As an example
here we mention the recently studied tripartite entanglement of
seven qubits used in connection with the $E_{7(7)}$ symmetric
black hole entropy formula regarded as an entanglement
measure\cite{Ferrara,Levay}. This entanglement measure is again
just the quartic invariant $q(x)$ (see Eq.(\ref{q})) for a
Freudenthal triple system however, now it is based on the cubic
Jordan algebra of $3\times 3$ Hermitian matrices with elements
taken from split octonions. There is a truncation of this formula
related to cubic Jordan algebras based on the split quaternions,
with the correponding entanglement regarded as a truncation of
this unusual type of tripartite entanglement. Our results on
three-fermion systems fit naturally into this scheme. Our cubic
Jordan algebra $M(3, {\bf C})$ can be shown to be isomorphic to the
one based on the split complex numbers i.e. the
binarions\cite{Mc}. Moreover, we have already seen that
three-qubit systems can be regarded as a convenient truncation of
this case. In summary all these cases of special entangled systems
fit nicely into the theory of Freudenthal triple systems based on
cubic Jordan algebras over split division algebras of complex
numbers, quaternions and octonions. The question is whether the
highly special entangled systems arising in the black hole context
have any relevance to quantum information theory.

We would like to emphasize that the nice results that we can
obtain for three-fermion systems with six single particle states
rest on the relationship between these systems and the very
special structure of Freudenthal systems. So for the general case
of multipartite fermionic entanglement these structures are not
useful. Hence the identification of different types of genuine
multipartite fermionic entanglement via suitable measures remains
a basic challenge. However, we would like to point out that as far
as the problem of separability for fermionic systems with
arbitrary number of constituents and single particle states is
concerned  Pl\"ucker relations provide a sufficient and necessary
condition of separability\cite{Harris,Hodge}. Hence if we have a
$k$-fermionic state with $n$ single particle states we can define
the $k$-form
\begin{equation}
P=\frac{1}{k!}\sum_{a_1a_2\dots a_k=1}^n P_{a_1a_2\dots a_k}e^{a_1}\wedge
e^{a_2}\wedge\dots\wedge e^{a_k}\in \bigwedge^{k}{\bf C}^n,
\label{genferm}
\end{equation}
\noindent Then as usual we call $P$ {\it separable} iff
$P={\omega}_1\wedge{\omega}_2\wedge\dots\wedge {\omega}_k$ for
some ${\omega}_j\in {\bf C}^n$. The sufficient and necessary
condition for this to happen is
\begin{equation}
\Pi_{{\cal A}, {\cal
B}}(P)=\sum_{j=1}^{k+1}(-1)^{j-1}P_{a_1a_2\dots
a_{k-1}b_j}P_{b_1b_2\dots b_{k+1}\hat{b}_j}=0, \label{general}
\end{equation}
\noindent where ${\cal A}=\{a_1,a_2,\dots ,a_{k-1}\}$ and ${\cal
B}=\{b_1,b_2,\dots ,b_{k+1}\}$ are $k-1$ and $k+1$ element subsets
of the set $\{1,2,\dots, n\}$, and where the number $\hat{b}_j$
has to be omitted . It is known\cite{Kasman} that we do not need
to consider all the choices of indices ${\cal A}$ and ${\cal B}$.
 We have to merely consider  the elements in ${\cal A}$ and ${\cal B}$ in increasing order. If
${\cal A}$ is contained in ${\cal B}$ the Pl\"ucker relations are
identically zero. If we have a single element $a\in {\cal A}$
which is {\it not} lying in the intersection of ${\cal A}$ and
${\cal B}$ we can demand that $a<b$ for all $b\in {\cal B}$ not in
the intersection. It is then calculated that the number of such
subsets is
\begin{equation}
\kappa=\frac{1}{4}+\sum_{m=1}^Ma_m,\quad
a_m=\frac{n!}{(m+1)!(m+3)!(k-m-2)!(n-k-m-2)!},\label{kombi}
\end{equation}
\noindent where $M={\rm min}\{k,n-k\}$. So ${\kappa}$ gives the
number of relations to be  checked. Moreover, it was also
shown\cite{Kasman} that one can construct a certain finite set of
maps mapping the original $k$-fermion state with $n$ single
particle states to a finite number of two-fermion states with four
single particle ones. It was shown that the separability of the
$k$-fermion state is in some sense equivalent  to the separability
of the corresponding two-fermion states. Recall the simplicity of
the Pl\"ucker relation in this case (see Eq. (\ref{plucky})). So
the measure ${\eta}$ based on this relation as far as the
separability of $k$-fermion states is concerned in some sense
universal.

There are a lot of further interesting questions to be addressed.
For example it is well-known that the three-tangle ${\tau}_{123}$
is an entanglement monotone. What about our newly introduced
quantity ${\cal T}_{123}$? There is also the question whether we
can generalize our tripartite measure via the usual convex roof
construction to obtain a corresponding mixed state measure.
Moreover, in the original three-qubit setup ${\tau}_{123}$ plays
the role of the residual tangle in the Coffman-Kundu-Wootters
relations\cite{Kundu} of distributed entanglement. Can these
relations be generalized in some sense? We can hopefully address
these interesting questions in future works.

\section{Acknowledgment}
One of us (P. L.) would like to thank Professor Werner Scheid for
the warm hospitality at the Department of Theoretical Physics at
the Justus Liebig University of Giessen where a part of this work
was completed.
Financial support from the Orsz\'agos Tudom\'anyos Kutat\'asi Alap (OTKA)
(Grants No. T047035, T047041, and T038191) is gratefully acknowledged.

\end{document}